\documentclass{aastex63}
\usepackage{graphicx}

\newcommand\rxj{ RXJ$1131$}
\newcommand{\code}[1]{\texttt{#1}}

\newcommand{\feka}{\hbox{Fe\,K$\alpha$}}

\newcommand{\msun}{\hbox{${M}_{\odot}$}}

\newcommand{\be}{\begin{equation}}
\newcommand{\ee}{\end{equation}}
\newcommand{\ba}{\begin{eqnarray}}
\newcommand{\ea}{\end{eqnarray}}

\newcommand{\xmm}{\emph{XMM-Newton}}

\newcommand{\swift}{\emph{Swift}}

\newcommand{\nustar}{\emph{NuSTAR}}

\newcommand{\simgt}{\lower 2pt \hbox{$\, \buildrel {\scriptstyle >}\over {\scriptstyle\sim}\,$}}
\newcommand{\simlt}{\lower 2pt \hbox{$\, \buildrel {\scriptstyle <}\over {\scriptstyle\sim}\,$}}
\newcommand{\ls}{\lower 2pt \hbox{$\;\scriptscriptstyle \buildrel<\over\sim\;$}}
\newcommand{\gs}{\lower 2pt \hbox{$\;\scriptscriptstyle \buildrel>\over\sim\;$}}

\begin{document}

\title{NuSTAR Hard X-ray Monitoring of Gravitationally Lensed Quasar RX\,J1131$-$1231}

\author[0000-0002-1650-7936]{Cora A. DeFrancesco}
\affiliation{Homer L. Dodge Department of Physics and Astronomy, The
  University of Oklahoma, Norman, OK, USA}
\author[0000-0001-9203-2808]{Xinyu Dai}
\affiliation{Homer L. Dodge Department of Physics and Astronomy, The
  University of Oklahoma, Norman, OK, USA}
\author{Mark Mitchell}
\affiliation{Homer L. Dodge Department of Physics and Astronomy, The
  University of Oklahoma, Norman, OK, USA}
\author{Abderahmen Zoghbi}
\affiliation{Department of Astronomy, University of Maryland, College Park, MD 20742}
\affiliation{HEASARC, Code 6601, NASA/GSFC, Greenbelt, MD 20771}
\affiliation{CRESST II, NASA Goddard Space Flight Center, Greenbelt, MD 20771}
\author{Christopher W. Morgan} 
\affiliation{Physics Department, United States Naval Academy, Annapolis, MD 21403, USA}

\begin{abstract}

The X-ray emission from active galactic nuclei (AGN) is believed to come from a combination of inverse Compton scattering of photons from the accretion disk and reprocessing of the direct X-ray emission by reflection. 
We present hard (10-80 keV) and soft (0.5-8 keV) X-ray monitoring of a gravitationally lensed quasar RX\,J1131$-$1231 (hereafter\rxj{}) with NuSTAR, Swift, and XMM-Newton between 10 June 2016 and 30 November 2020.  
Comparing the amplitude of quasar microlensing variability at the hard and soft bands allows a size comparison, where larger sources lead to smaller microlensing variability.
During the period between 6 June 2018 and 30 November 2020, where both the hard and soft light curves are available, the hard and soft bands varied by factors of 3.7 and 5.5, respectively, with rms variability of $0.40 \pm 0.05$ and $0.57 \pm 0.02$. 
Both the variability amplitude and rms are moderately smaller for the hard X-ray emission, indicating that the hard X-ray emission is moderately larger than the soft X-ray emission region.
We found the reflection fraction from seven joint hard and soft X-ray monitoring epochs is effectively consistent with a constant with low significance variability. 
After decomposing the total X-ray flux into direct and reprocessed components, we find a smaller variability amplitude for the reprocessed flux compared to the direct emission.
The power-law cutoff energy is constrained at 96$^{+47}_{-24}$~keV, which position the system in the allowable parameter space due to the pair production limit.

\end{abstract}


\section{Introduction} \label{sec:intro}

Quasars spectral energy distributions range from radio to X-ray bands \citep[e.g.,][]{1994ApJS...95....1E, 2011ApJS..196....2S, 2020MNRAS.492..315G}, with jetted ones extending to the gamma-ray regime \citep[e.g.,][]{2020ApJS..247...33A, 2022ApJS..262...18Y}. 
This diversity of features comes from the aggregated emission of many different physical components in the central engine \citep[e.g.,][]{1993ARA&A..31..473A, 1995PASP..107..803U}. In particular, the X-ray spectrum of quasars is composed of a direct X-ray continuum, a soft excess, and a Compton hump \citep[e.g.,][]{1995MNRAS.277L..11F, 1997ApJ...477..602N, 2011ApJ...727...19F}. The direct X-ray flux is generated by inverse Compton scattering, where ultra-violet photons from the accretion disk interact with relativistic electrons from the corona, causing the photons to gain energy and emit a power-law X-ray continuum \citep{Haardt_Maraschi_1991, Haardt_Maraschi_1993}. 
This X-ray continuum will irradiate the cold, optically thick material, and 
Compton scattering and photoelectric absorption will produce a reflection spectrum with associated metal fluorescent lines.
The most prominent features of the reflection component are the FeK$\alpha$ line (6.4 keV) due to the high cosmic iron abundance and fluorescent yield and further enhanced by the fact that the reflection hump peaks at $\sim20$~keV (see \citet{Reynolds_Nowak_2003} for a review).  

The spatial origin of the reflection region has been debated in the literature \citep[e.g.,][]{2022A&A...664A..46A, 2022ApJ...936...66M}, on whether the reflection occurs in a region immediately around the supermassive black hole, accretion disk, disk wind, or torus. 
The FeK$\alpha$ line profile provides a first clue to the reflection location.  Since a narrow feature is commonly observed in the X-ray spectrum of AGN \citep[e.g.,][]{2001ApJ...554..216K, 2001ApJ...546..759Y, 2005ApJ...618L..83Z}, remote reflections, e.g., off the outer regions of the accretion disk or torus, should contribute to a portion of the reflection flux. 
For some nearby Compton-thick AGN, spatially resolved reflection component has been measured in a range of a few tens to several hundreds of parsecs around the central engine \citep[e.g.,][]{2017ApJ...842L...4F, 2019PASJ...71...68K, 2020ApJ...900..164M, 2021ApJ...908..156Y}.
In addition, a broad and skewed relativistic line component is observed in many high signal-to-noise ratio spectra of AGN (e.g., MCG--6-30-5, \citet{Tanaka_1995}; NGC 3516, \citet{Nandra_1997}), suggesting that the reflection in a region immediately around the supermassive black hole is present in many AGN \citep{Fabian_1989, Laor_1991}.
Light bending models were introduced \citep{Miniutti_Fabian_2004}, where the main X-ray emission is from a very compact source immediately above the black hole.
Because of the strong gravitational light bending, the majority of the X-ray emission will return and irradiate the accretion disk and only a smaller fraction of the direct continuum will escape.
If the source has vertical motion, it will result in a time varying partition between 
the direct and returned emission, and the fractional change of the direct component will be much larger than that of the reflected due to the different strengths of the light bending effect.
These models require a very compact power-law source, which is confirmed by recent quasar microlensing \citep[e.g.,][]{2008ApJ...689..755M, 2012ApJ...756...52M, Dai_2010, Mosquera_2013, 2016AN....337..356C,2017ApJ...836..206G} and X-ray reverberation mapping methods (e.g. \citet{Reis_Miller_2013}). 
Separating the broad iron line from the underlying continuum may be complicated sometimes by the spectral curvature introduced by the partially ionized, and possibly non-uniform, absorbers \citep[e.g.,][]{Reeves_2018}.
It is also certainly possible that the reflection component is produced by a combination of heterogeneous sources.

A key to resolving this issue is to measure the relative size of the reflection component with respect to the (power law emitting) X-ray continuum to constrain the majority of the reflection region.
Quasar microlensing has become a powerful tool to measure different components of AGN central engines that cannot be spatially resolved by current telescopes, with equivalent nano-arcsec resolutions \citep{2007ASPC..371...43K, 2019arXiv190412967M}. 
Quasar microlensing \citep{1989AJ.....98.1989I, 1989A&A...215....1V} is induced by stars in the lensing galaxy close to the gravitationally lensed images yielding extra magnifications or demagnifications to those images.  As the source, lens, and observers move relatively, quasar microlensing will produce uncorrelated variability between the lensed images. Furthermore, over the relatively large energy band considered in our study, there is the possibility of two components that could in principle vary independently. 
An important scale in these studies is the Einstein ring size in the source plane for a typical microlens, where sources smaller than this scale are sensitive to microlensing effects, and for a 1\msun\ star, the typical Einstein ring size is slightly smaller than the estimated broad line region sizes \citep[e.g.,][]{Mosquera_2011}. 
Thus, quasar microlensing has been used to constrain quasar central engine components from the broad line to the X-ray regions.  
Recent studies also show that planet mass objects in the lens galaxy will contribute significantly to the X-ray microlensing signal, because their Einstein ring scales match the compact emission regions immediately around the supermassive black hole \citep{2018ApJ...853L..27D, 2019ApJ...885...77B, 2020ApJ...896..111G}.
Since quasar microlensing has successfully constrained the relativistic \feka\ region to be 2--8 $r_g$ in a sample of gravitationally lensed quasars \citep{Dai_2019}, this implies that the reflection hump should have a similar size. 
Microlensing size measurements of the reflection region will provide an independent test to the current paradigm of relativistic reflection model, where the corona is compact with 10\,$r_g$ and the reflection region is dominated by relativistic effects including light bending.

RX\,J1131$-$1231 (hereafter\rxj{}) is a quadruply lensed quasar system with the lens and source redshift at $z_l = 0.295$ and $z_s = 0.658$, respectively \citep{2003A&A...406L..43S}, where
the central black hole mass of the lensed quasar is estimated to be $M_{BH} = (1.3\pm0.3) \times 10^{8} \msun$ \citep{Dai_2010}.
The system shows large quasar microlensing variability based on 38 Chandra monitoring observations over a decade, where the image resolved flux ratios exhibit large variations \citep{Chartas_2017}.  
It is also the brightest radio-quiet lensed quasar in the X-ray band  \citep{Jackson_2015}.  These properties make \rxj\ an ideal target to monitor with \nustar\ to constrain the relative size of the reflection hump compared to the direct X-ray continuum, and here we present the analysis of these \nustar\ monitoring observations paired with soft X-ray observations by \swift\ or \xmm.
The paper is organized as follows. We describe the observations and data analysis procedures in Section~\ref{sec:data}, spectral analysis in Section~\ref{sec:spec}, and present the analysis results and discussion in Section~\ref{sec:dis}.  We assumed the set of cosmological parameters $\Omega_{m} = 0.3$, $\Omega_{\Lambda} = 0.7$, and $H_0 = 70$~km~s$^{-1}$~Mpc$^{-1}$ in this paper.

\section{X-Ray Observations and Data Analysis \label{sec:data}} 

We have performed six pairs of joint observations with the Nuclear Spectroscopic Telescope Array (\textit{NuSTAR}, \citet{Harrison_2013}) and the Neil Gehrels Swift Observatory (\textit{Swift}, \citep{Burrows_2005}) to monitor the hard (10-80 keV) and soft (0.5-8 keV) X-ray variability of \rxj{} between 6 June 2018 and 30 November 2020. The \textit{Swift} observations averaged an exposure time of 1.49 ks, and the \textit{NuSTAR} observations 30 ks. We also analyzed a pair of archival joint observations from the X-ray Multi-Mirror Mission (\textit{XMM-Newton}, \citet{Jansen_2001}) with 97.7 ks exposure and \textit{NuSTAR} with 77.4 ks exposure on 6 June 2018.
We additionally analyzed 32 \textit{Swift} observations between 20 June 2016 and 30 November 2020 besides the six paired with \textit{NuSTAR} observations. The log of observations is listed in Table \ref{table:obs_log}. 

\startlongtable
\begin{deluxetable*}{cccc}
\tabletypesize{\footnotesize}
\tablecolumns{4}
\tablewidth{0pt}
\tablecaption{Observation Log \label{table:obs_log}}

\tablehead{
\colhead{Observation ID}& \colhead{JD}  & \colhead{Exposure} & \colhead{Unabsorbed Flux} \\
\colhead{} & \colhead{} & \colhead{Time (s)} & \colhead{($10^{-12}$ ergs/cm$^2$/s)}}

\startdata
\tableline
\multicolumn{4}{c}{\textit{NuSTAR} FPMA Observations, $10.0-80.0$ keV} \\
\tableline
60401001002\tablenotemark{a} & 2458275.5 & 77500.0 & $ 6.87 ^{+ 0.13 }_{- 0.13 }$ \\
60502021002 & 2458683.5 & 20600.0 & $ 9.35 ^{+ 0.73 }_{- 0.67 }$ \\
60502021004 & 2458835.5 & 23600.0 & $ 4.62 ^{+ 0.52 }_{- 0.45 }$ \\
60502021006 & 2458876.5 & 20800.0 & $ 3.58 ^{+ 0.54 }_{- 0.34 }$ \\
60502021008 & 2458974.5 & 22600.0 & $ 5.35 ^{+ 0.57 }_{- 0.37 }$ \\
60502021010 & 2459015.5 & 22900.0 & $ 4.29 ^{+ 0.44 }_{- 0.40 }$ \\
60502021012 & 2459183.5 & 21200.0 & $ 2.38 ^{+ 0.44 }_{- 0.37 }$ \\
\tableline
\multicolumn{4}{c}{\textit{NuSTAR} FPMB Observations, $10.0-80.0$ keV} \\
\tableline
60401001002\tablenotemark{a} & 2458275.5 & 77300.0 & $ 7.99 ^{+ 0.12 }_{- 0.12 }$ \\
60502021002 & 2458683.5 & 20500.0 & $ 9.77 ^{+ 0.75 }_{- 0.70 }$ \\
60502021004 & 2458835.5 & 23400.0 & $ 5.18 ^{+ 0.58 }_{- 0.50 }$ \\
60502021006 & 2458876.5 & 20600.0 & $ 3.86 ^{+ 0.57 }_{- 0.36 }$ \\
60502021008 & 2458974.5 & 22400.0 & $ 5.84 ^{+ 0.61 }_{- 0.39 }$ \\
60502021010 & 2459015.5 & 22800.0 & $ 5.21 ^{+ 0.53 }_{- 0.48 }$ \\
60502021012 & 2459183.5 & 21000.0 & $ 2.84 ^{+ 0.54 }_{- 0.45 }$ \\
\tableline
\multicolumn{4}{c}{\textit{XMM-Newton} Observations, $0.5-8$ keV} \\
\tableline
0820830101\tablenotemark{a} & 2458275.5 & 97684.8 & $ 5.71 ^{+ 0.03 }_{- 0.02 }$ \\
\tableline
\multicolumn{4}{c}{\textit{Swift} Observations, $0.5-8$ keV} \\
\hline
34575001 & 2457518.71624 & 2230.0 & $ 1.88 ^{+ 0.15 }_{- 0.27 }$ \\
34575002 & 2457526.42324 & 1450.0 & \nodata \\
34575003 & 2457533.41341 & 1960.0 & $ 2.06 ^{+ 0.18 }_{- 0.32 }$ \\
34575004 & 2457541.44889 & 87.4   & \nodata \\
34575005 & 2457546.76301 & 1620.0 & $ 1.52 ^{+ 0.20 }_{- 0.19 }$ \\
34575006 & 2457554.87094 & 602.0 & \nodata \\
34575007 & 2457562.31264 & 564.0 & \nodata \\
34575008 & 2457563.57753 & 1220.0 & $ 2.20 ^{+ 0.33 }_{- 0.44 }$ \\
34575009 & 2457568.90892 & 1060.0 & $ 2.36 ^{+ 0.34 }_{- 0.37 }$ \\
34575010 & 2457699.38222 & 1960.0 & $ 2.82 ^{+ 0.18 }_{- 0.26 }$ \\
34575011 & 2457712.58699 & 1920.0 & $ 1.96 ^{+ 0.03 }_{- 0.33 }$ \\
34575012 & 2457727.01353 & 1970.0 & $ 2.60 ^{+ 0.26 }_{- 0.29 }$ \\
34575013 & 2457741.17628 & 1030.0 & $ 2.85 ^{+ 0.40 }_{- 0.42 }$ \\
34575014 & 2457746.47229 & 1040.0 & $ 2.00 ^{+ 0.10 }_{- 0.46 }$ \\
34575015 & 2457755.17853 & 1870.0 & $ 2.64 ^{+ 0.31 }_{- 0.29 }$ \\
34575016 & 2458542.80707 & 2000.0 & $ 6.97 ^{+ 0.18 }_{- 0.51 }$ \\
34575017 & 2458549.526 & 1440.0 & $ 4.72 ^{+ 0.35 }_{- 0.49 }$ \\
34575018 & 2458556.68231 & 1010.0 & $ 3.63 ^{+ 0.46 }_{- 0.42 }$ \\
34575019 & 2458561.7308 & 926.0 & $ 5.80 ^{+ 0.20 }_{- 0.86 }$ \\
34575020 & 2458565.71318 & 1900.0 & $ 5.79 ^{+ 0.37 }_{- 0.34 }$ \\
34575021 & 2458571.38119 & 290.0 & $ 6.56 ^{+ 0.34 }_{- 2.21 }$ \\
34575022 & 2458584.84601 & 2000.0 & $ 11.35 ^{+ 0.25 }_{- 0.70 }$ \\
34575023 & 2458592.08391 & 1930.0 & $ 8.34 ^{+ 0.42 }_{- 0.44 }$ \\
34575024 & 2458598.58789 & 1980.0 & $ 9.65 ^{+ 0.21 }_{- 0.740 }$ \\
34575025 & 2458606.0412 & 1500.0 & $ 8.03 ^{+ 0.24 }_{- 0.71 }$ \\
34575026 & 2458613.32997 & 759.0 & $ 7.62 ^{+ 0.54 }_{- 0.89 }$ \\
34575027 & 2458619.96542 & 1730.0 & $ 5.48 ^{+ 0.36 }_{- 0.29 }$ \\
34575028 & 2458626.53948 & 1790.0 & $ 4.01 ^{+ 0.24 }_{- 0.24 }$ \\
34575029 & 2458633.77734 & 2070.0 & $ 6.56 ^{+ 0.38 }_{- 0.42 }$ \\
88895001 & 2458684.37758 & 1940.0 & $ 6.19 ^{+ 0.35 }_{- 0.31 }$ \\
88895002 & 2458828.46554 & 1350.0 & $ 4.37 ^{+ 0.41 }_{- 0.39 }$ \\
88895003 & 2458835.78503 & 986.0 & $ 2.73 ^{+ 0.13 }_{- 0.75 }$ \\
88895004 & 2458877.13365 & 697.0 & $ 2.05 ^{+ 0.36 }_{- 0.83 }$ \\
88895005 & 2458880.18729 & 1200.0 & $ 2.90 ^{+ 0.21 }_{- 0.59 }$ \\
88895006 & 2458975.53266 & 1980.0 & $ 3.70 ^{+ 0.15 }_{- 0.41 }$ \\
88895007 & 2459016.04402 & 2310.0 & $ 3.31 ^{+ 0.27 }_{- 0.24 }$ \\
88895008 & 2459016.51149 & 435.0 & $ 2.42 ^{+ 0.37 }_{- 0.82 }$ \\
88895008 & 2459184.14302 & 1910.0 & $ 1.22 ^{+ 0.12 }_{- 0.17 }$ \\
\tableline
\enddata

\tablenotetext{a}{Archival observation reanalyzed in this paper.}


\end{deluxetable*}

We reduced the data using \code{HEAsoft} version 6.25 \citep{Heasarc_2014}. The \textit{NuSTAR} observations from Focal Plane Modules A and B (FPMA, FPMB) were processed with \code{NUSTARDAS} using CALDB version 1.0.2. We reprocessed the data using the \textit{NuSTAR} data analysis pipeline \code{nupipeline}. We used the \code{SCIENCE} observation mode and did not filter for SAA after experimentation with different filtering methods showed consistent results with unfiltered data. The Stage 2 event files were used to further select source and background regions in accordance with the recommendations from the analysis guide. The circular source region was centered on the coordinates of \rxj{} with a radius of 61.5''. The background region had a radius of 196.8'' and overlapped two quadrants. Default grade selection of grades 0--26 was used.
Because of the slight differences in the FPMA and FPMB telescopes, the spectra were not co-added and instead were simultaneously fit in the spectral analysis. Ancillary response files were generated for each spectrum as part of Stage 3 processing with \code{nupipeline} through the \code{numkarf} subroutine.

The \textit{Swift} data reduction was performed using \code{XSelect}. We used the response matrix file (RMF) for photon counting mode for default grades 0--12. Source and background region files were created. The circular source region was centered on the coordinates of \rxj{} with a radius of 47.2'', and the background region was placed significantly away from the source and had a radius of 295''. 
The ancillary response file (ARF) was made for each observation using \code{xrtmkarf}. 
We binned the spectra by 15 counts to standardize the SNR for each bin. Some observations did not have enough counts to support this binning scheme; they were binned by 7 counts (see note in Table \ref{table:obs_log}). We performed the spectral fitting in XSPEC \footnote{v.12.11 (https://heasarc.gsfc.nasa.gov/xanadu/xspec/manual/node1.html)}.

The \textit{XMM-Newton} EPIC data for observation 0820830101 were reduced using \code{epchain} in SAS (xmmsas-20201028-0905-19.0.0). Source and background photons with standard grades are extracted from circular regions of 50'' radius centered on and away from the location of the source respectively. The response and area files were produced using the \code{rmfgen} and \code{arfgen} tools respectively.

\section{Spectral Analysis \label{sec:spec}}

\subsection{\textit{Swift} Spectra}

\begin{figure}
    \centering
    \includegraphics[width=\linewidth]{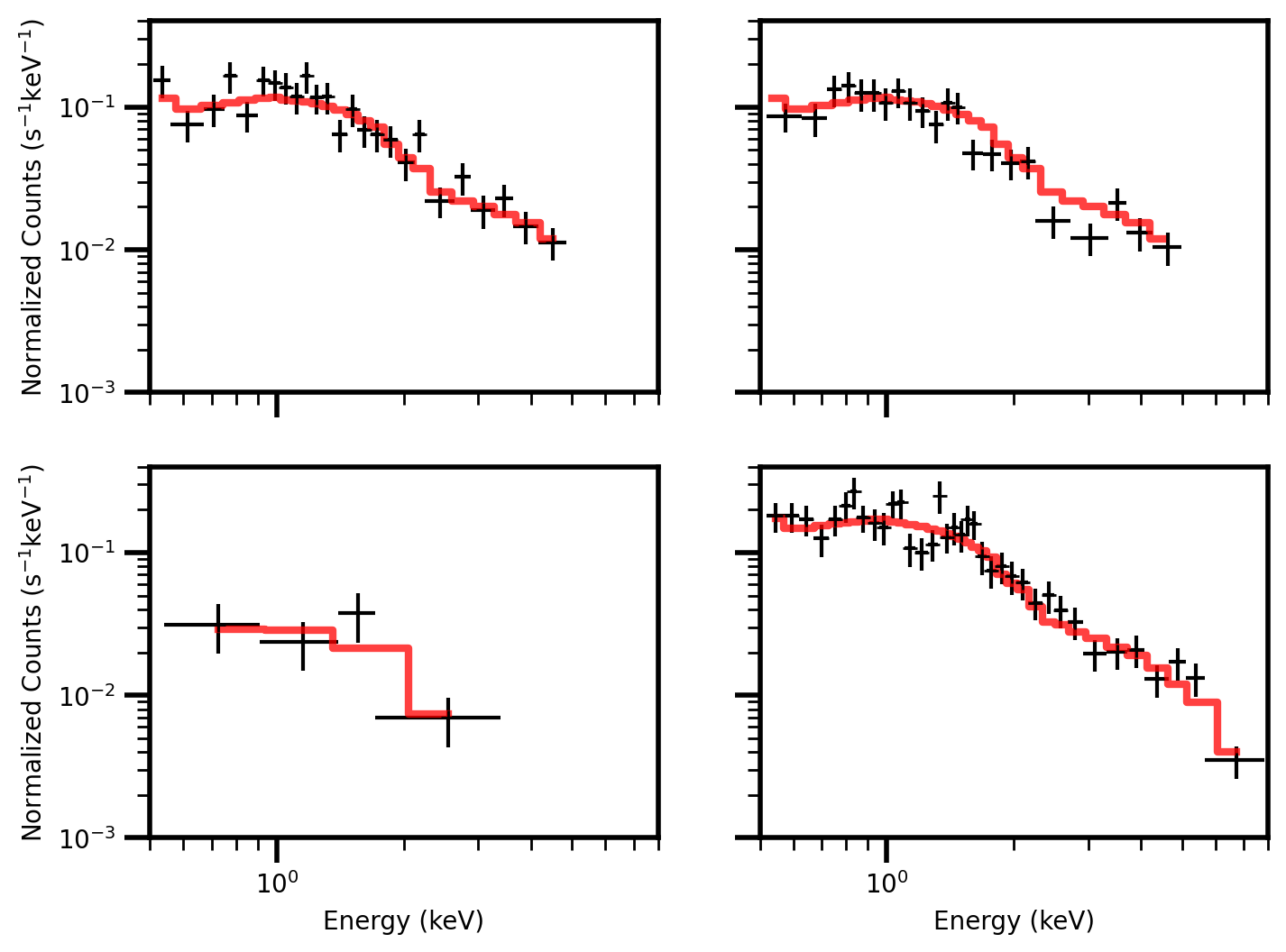}
    \caption{Example \textit{Swift} spectra. Top left and top right: Typical SNR for the \swift\ spectra (19 and 18, respectively). Bottom left: An example of low SNR spectrum (4.8) excluded from analysis. Bottom right: An example of high SNR spectrum (24).}
    \label{fig:sw_examps}
\end{figure}

We fit the 38 \swift\ spectra in the energy range $0.5 - 8$ keV, ignoring the rest as recommended by the manual. We used $\chi^2$ statistics for spectra binned by a minimum of 15 counts and C-statistics for those binned by 7 counts.

Because of the low SNR of these \textit{Swift} spectra, with a median net count of 185, we can only fit these spectra with a simple powerlaw (\code{zpowerlw}) modified by Galactic (\code{tbabs}) and intrinsic absorption (\code{ztbabs}). We froze the Galactic neutral hydrogen column densities at $2.68 \times 10^{20} \text{atoms/cm}^2$, which we obtained from the HEASARC online tool \footnote{https://heasarc.gsfc.nasa.gov/cgi-bin/Tools/w3nh/w3nh.pl}. We found during the fitting process that the intrinsic absorption component is consistent with zero and therefore we dropped it in the subsequent analysis. This is consistent with the results from much higher SNR Chandra spectra \citep{Chartas_2017}, which also constrained the intrinsic absorption to be zero. Several example spectra are shown in Figure \ref{fig:sw_examps}.

Thus for each \textit{Swift} spectrum, we are able to constrain the photon index and normalization. The measured photon indices range from 1.21 to 2.71 keV with a median of $\Gamma=1.69$, which is consistent with the median of the Chandra spectra \citep{Chartas_2017}. The uncertainties of the photon index are typically between 10 to 20 percent set by the typical SNR of these spectra.  Finally, we set the photon index to the median value and refit the model to measure the unabsorbed model flux in the soft X-ray band from the 38 \swift\ spectra (Table \ref{table:obs_log}). 

\subsection{Joint Soft and Hard X-ray Spectral Analysis}

We analyzed the six \textit{Swift}/\textit{NuSTAR} and one \textit{XMM-Newton}/\textit{NuSTAR} spectral pairs jointly to constrain the reflection fraction of \rxj{} over the soft and hard X-ray bands. The spectral analysis was performed in 0.5 -- 8.0, 10.0 -- 80.0, and 0.5 -- 8.0 keV bands for \textit{Swift}, \textit{NuSTAR}, and \textit{XMM-Newton}, respectively.

\begin{figure}
    \centering
    \includegraphics[width=\linewidth]{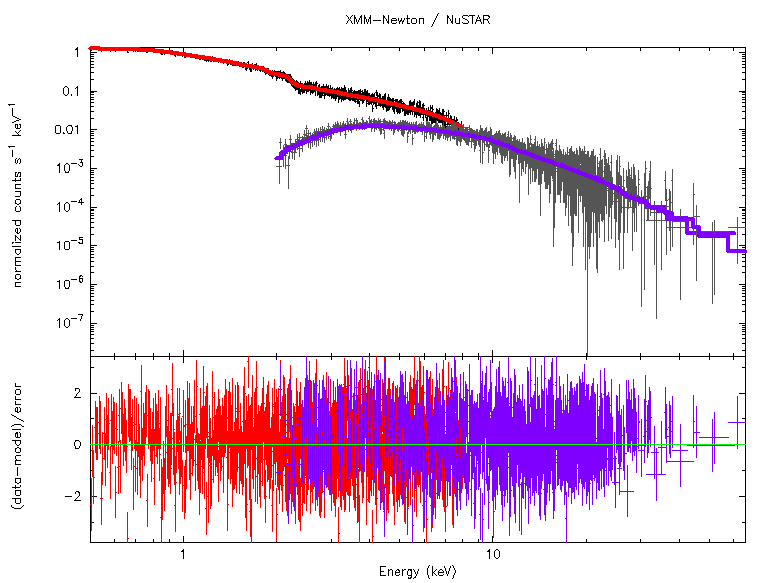}
    \caption{Top: Combined \textit{XMM-Newton} (low energy) / \textit{NuSTAR} (high energy) fit. The model is represented by the solid red (purple) line overlaying the \textit{XMM-Newton} (\textit{NuSTAR}) data. \textit{NuSTAR} data is shown in two data sets coming from the two focal plane modules FPMA and FPMB. Bottom: Residuals of the fit.}
    \label{fig:xmm_nu_fit}
\end{figure}

We first focused on the high SNR \textit{XMM-Newton/NuSTAR} spectral pair to guide our spectral analysis process. We modeled the \textit{XMM-Newton}/\textit{NuSTAR} pair as an exponentially cut off power law reflected from neutral material (\code{pexrav}) or ionized material (\code{pexriv}) \citep{Magdziarz_1995} modified by Galactic absorption (\code{tbabs}). We binned the \textit{NuSTAR} spectra by SNR of 1 which yielded the tightest constraints on model parameters based on our experiments of a range of binning schemes. The XMM spectrum was binned to have a minimum of 25 counts per bin with a minimum SNR of 6. We froze the model parameters for redshift, Galactic nH, metal and iron abundances (at solar abundances), and inclination (at 30$^\circ$, consistent with Seyfert 1 AGN). The values of the fit parameters were tied across all three spectra. The neutral reflection model also included a Gaussian centered at 0.7 keV to account for the soft excess component \citep{Reis_2014}. 
The final model fit for photon index, reflection fraction, normalization, and cutoff energy is listed in Table \ref{table:xmm_nu_res}, and the spectra and best-fit \code{pexrav} model were shown in Figure \ref{fig:xmm_nu_fit}.  
Both the neutral and ionized reflection models yield acceptable fits to the joint \textit{XMM-Newton}/\textit{NuSTAR} spectrum in terms of reduced $\chi^2$, with the reflection fraction and cutoff energy constrained at $r_{refl} = 0.22_{-0.10}^{+0.12}$, $0.34_{-0.10}^{+0.11}$ and $E_{cut} = 96_{-24}^{+47}$,   $107_{-27}^{+157}$, respectively, for the neutral and ionized reflection models.
Both models have similar uncertainties on the reflection fraction; however, the cutoff energy is less constrained in the ionized model on the upper range and the ionization parameter $\xi$ is poorly constrained for the ionized model. 

\begin{figure}
    \centering
    \includegraphics[width=\linewidth]{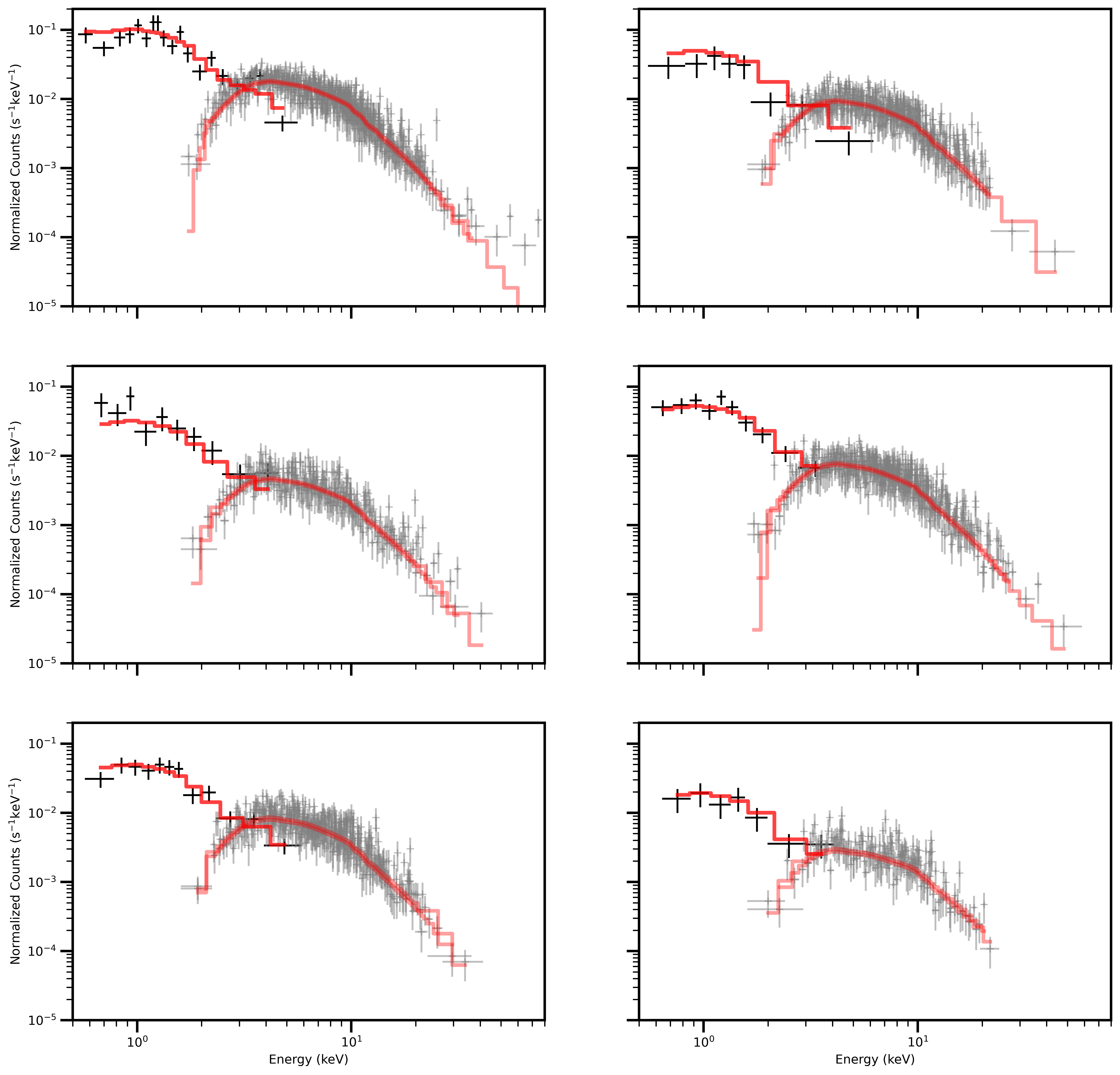}
    \caption{Model fits for the six \textit{Swift}/\textit{NuSTAR} paired observations. Soft X-ray \textit{Swift} spectra are in black, hard X-ray \textit{NuSTAR} spectra are in gray, and the model is in red.}
    \label{fig:comb_fits}
\end{figure}

Next, for the six lower SNR \textit{Swift}/\textit{NuSTAR} pairs, we used the characteristic photon index from the \textit{Swift} observations and the cutoff energy from the \textit{XMM-Newton}/\textit{NuSTAR} joint fit to constrain the reflection fraction. We used the simpler neutral reflection model and fixed most of the parameters as with the \textit{XMM-Newton}/\textit{NuSTAR} paired observation, but only varying the reflection fraction and the normalization of each pair. As before, the parameters for all spectra in an epoch were tied together. Our best constraints were derived from binning \textit{NuSTAR} spectra to a minimum SNR of 2 and \textit{Swift} spectra a minimum of 15 counts per bin. Our results are listed in Table \ref{table:refl_norm} and the temporal evolution of the reflection fraction is plotted in Figure \ref{fig:refl_curve}. Model fits for the six \textit{Swift}/\textit{NuSTAR} pairs are plotted in Figure \ref{fig:comb_fits}.

\startlongtable
\begin{deluxetable}{ccc}
\tabletypesize{\footnotesize}
\tablecolumns{3}
\tablewidth{0pt}
\tablecaption{\textit{XMM-Newton/NuSTAR} Fit Results \label{table:xmm_nu_res}}

\tablehead{
\colhead{Parameter}  & \colhead{Neutral Medium} & \colhead{Ionized Medium}}

\startdata
$\Gamma$ & $1.69_{-0.01}^{+0.01}$ & $1.70_{-0.01}^{+0.01}$\\
$E_{cut}$ $(\text{ keV})$ & $96_{-24}^{+47}$  & $102_{-23}^{+39}$ \\
$r_{refl}$ & $0.22_{-0.10}^{+0.12}$ & $0.23_{-0.09}^{+0.10}$\\
$\xi$ & \nodata & $336_{-247}^{+619}$ \\
norm ($10^{-4}$ photons/keV/cm$^2$/s)  & $8.2_{-0.1}^{+0.1}$ &  $8.19_{-0.07}^{+0.07}$\\
$\sigma$ (keV) & $0.25_{-0.03}^{+0.03}$ & $0.27_{-0.03}^{+0.03}$ \\
norm (gauss) ($10^{-4}$ photons/keV/cm$^2$/s)& $5.2_{-1.1}^{+1.2}$ & $5.4_{-1.2}^{+1.2}$\\
$\chi^2$ & 2004 & 1998\\
$\text{Reduced }$ $\chi^2$ & 1.04 & 1.03\\
\enddata
\end{deluxetable}

\startlongtable
\begin{deluxetable}{ccc}
\tabletypesize{\footnotesize}
\tablecolumns{3}
\tablewidth{0pt}
\tablecaption{Reflection fraction Fit Results \label{table:refl_norm}}

\tablehead{
\colhead{JD}  & \colhead{$r_{refl}$} & \colhead{Norm ($10^{-4}$ photons/keV/cm$^2$/s)}}

\startdata
2458275 & $ 0.22 ^{+ 0.12 }_{ -0.10 }$ & $ 8.20 ^{+ 0.01 }_{ -0.01 }$\\
2458683 & $ 0.14 ^{+ 0.06 }_{ -0.06 }$ & $ 12.17 ^{+ 0.28 }_{ -0.28 }$\\
2458835 & $ 0.22 ^{+ 0.09 }_{ -0.09 }$ & $ 6.01 ^{+ 0.2 }_{ -0.2 }$\\
2458876 & $ 0.27 ^{+ 0.13 }_{ -0.13 }$ & $ 3.93 ^{+ 0.18 }_{ -0.18 }$\\
2458974 & $ 0.22 ^{+ 0.09 }_{ -0.09 }$ & $ 6.17 ^{+ 0.2 }_{ -0.2 }$\\
2459015 & $ 0.11 ^{+ 0.09 }_{ -0.08 }$ & $ 5.83 ^{+ 0.18 }_{ -0.18 }$\\
2459183 & $ 0.65 ^{+ 0.21 }_{ -0.21 }$ & $ 2.25 ^{+ 0.14 }_{ -0.14 }$\\
\enddata
\end{deluxetable}

\section{Results and Discussion \label{sec:dis}}

\begin{figure}
    \centering
    \includegraphics[width=\linewidth]{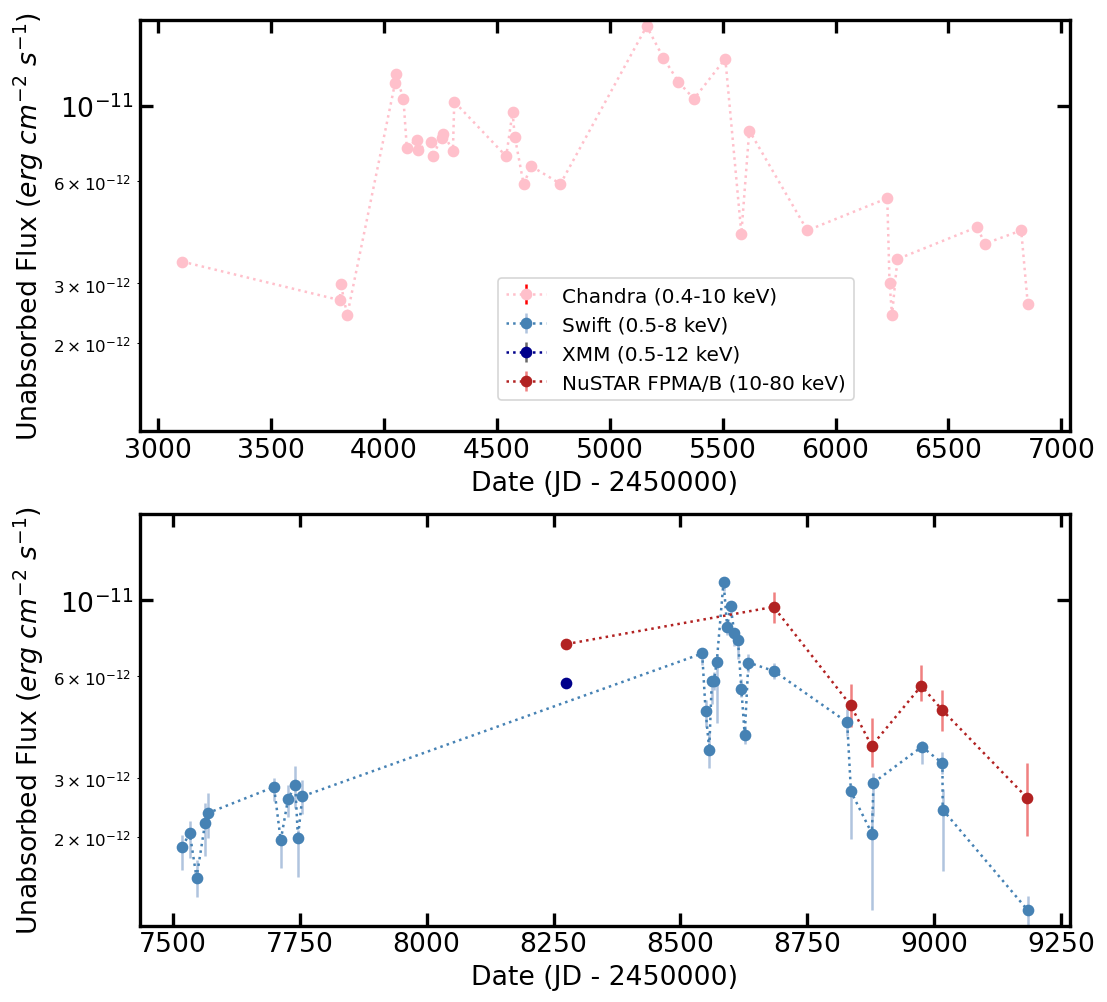}
    \caption{Unabsorbed flux light curve of all spectra from \textit{Nustar, Swift, XMM,} and \textit{Chandra}, where the \textit{Chandra} flux is from \citet{Chartas_2017}; all other data was described in this paper. The dotted lines are not physical and only used for the purpose of better reading the individual data points.}
    \label{fig:flux_lc}
\end{figure}

\begin{figure}
    \centering
    \begin{minipage}{0.49\textwidth}
        \centering
        \includegraphics[width=1.0\textwidth]{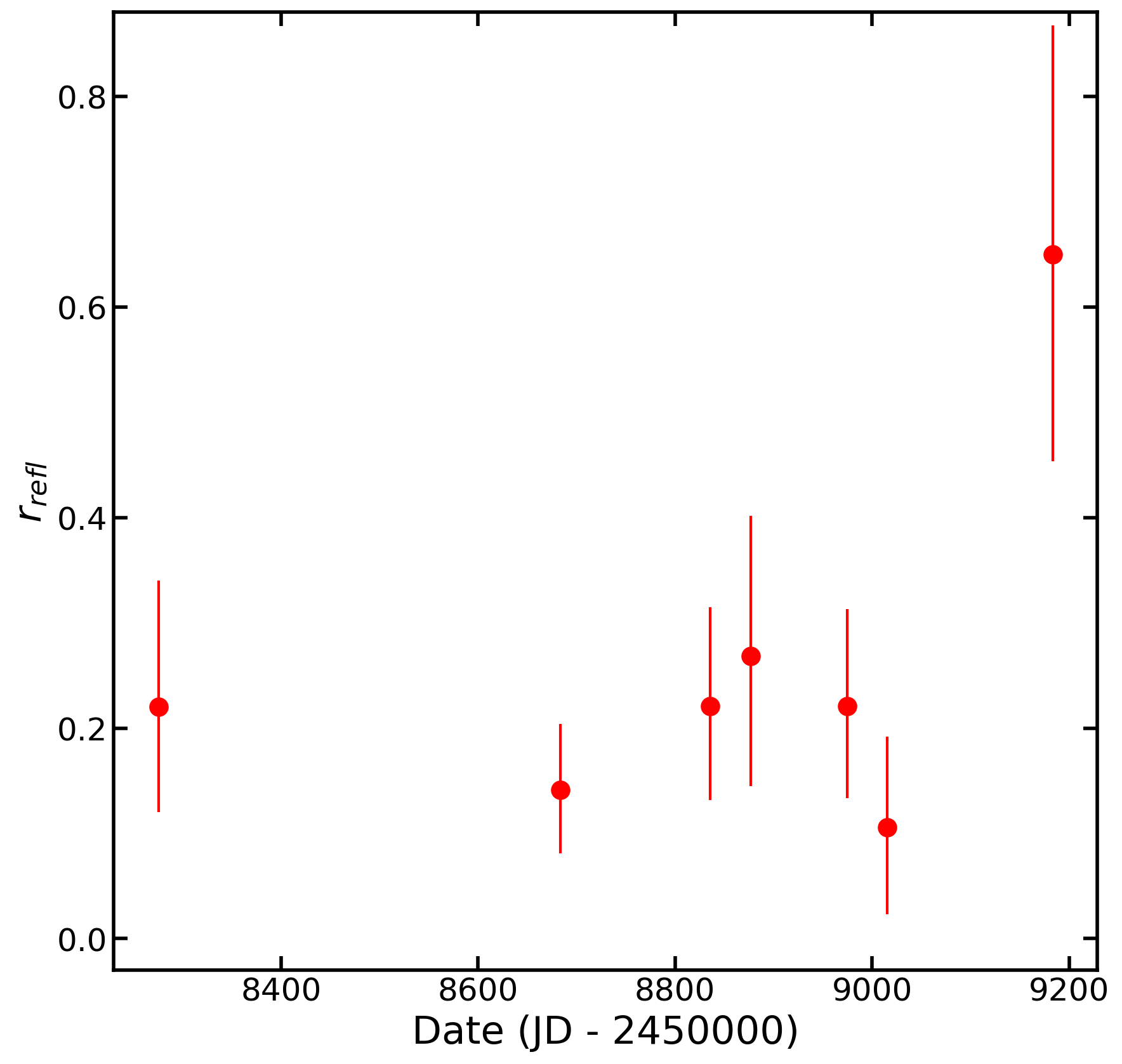} 
    \end{minipage}\hfill
    \begin{minipage}{0.49\textwidth}
        \centering
        \includegraphics[width=1.0\textwidth]{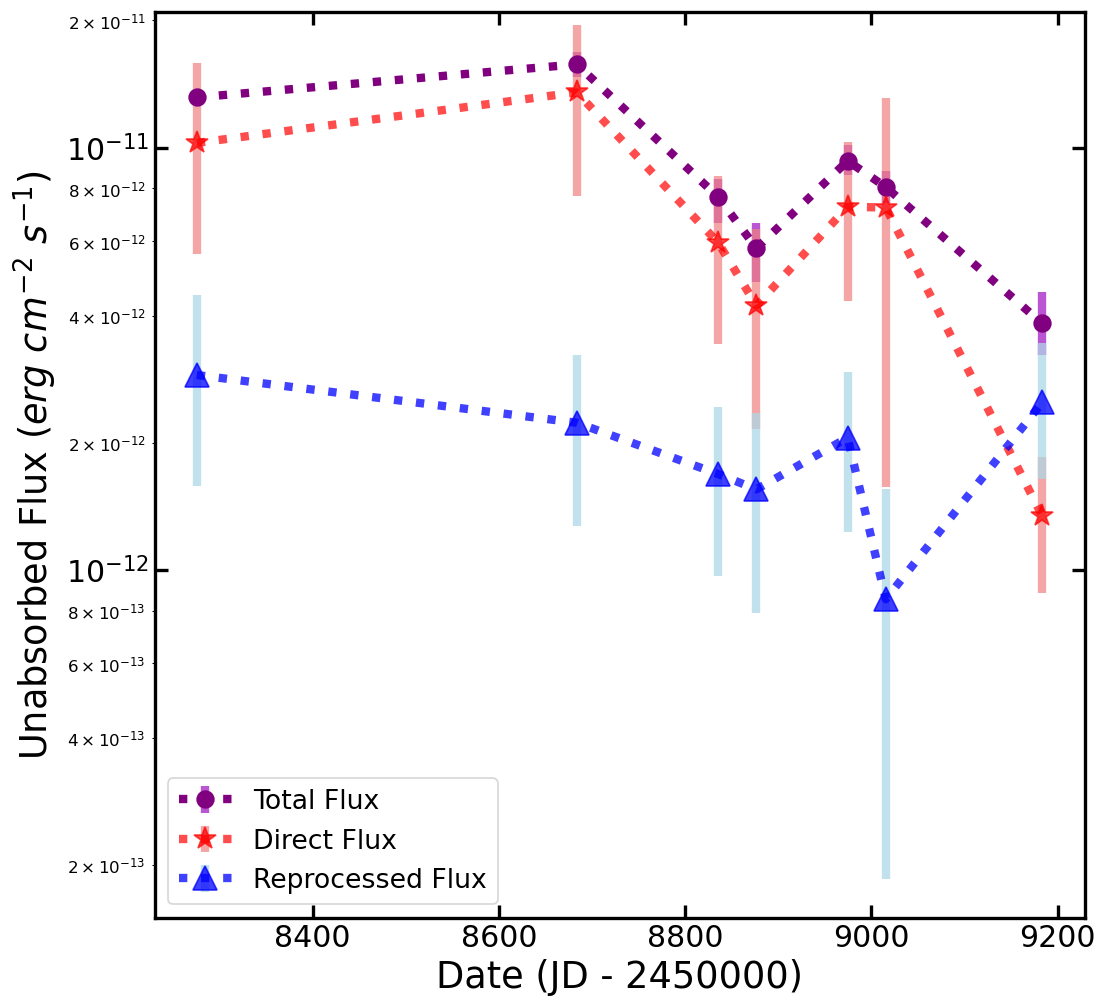} 
    \end{minipage}
    \caption{Left: Reflection fraction curve. The first point is from the archival \textit{XMM-Newton}-\textit{NuSTAR} pair; the rest are paired \textit{Swift}-\textit{NuSTAR} observations. Right: Change in total, direct and reprocessed emission over time.}
    \label{fig:refl_curve}
\end{figure}

\begin{figure}
    \centering
    \includegraphics[width=\linewidth]{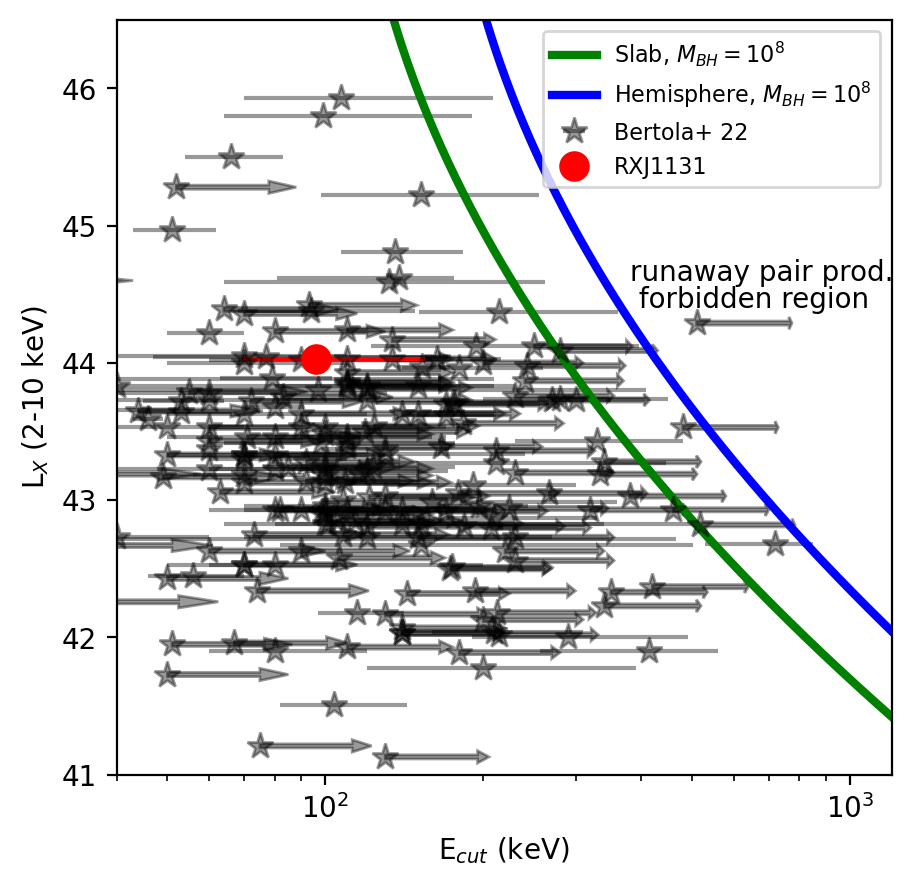}
    \caption{X-ray luminosity versus cutoff energy from a selection of quasars \citep{Bertola_2022}. The green and blue curves represent the boundary line for runaway pair production presented as a polynomial in \citet{Ricci_2018}. We assume an X-ray source size of $10 r_g$. Arrows in place of error bars represent a lower limit on the cutoff energy.}
    \label{fig:Lx_Ecut}
\end{figure}

We first present the hard and soft band light curves including both our new measurements from \textit{NuSTAR, Swift, XMM-Newton} and archival \textit{Chandra} measurements of the total flux from A, B, C, and D images \citep{Chartas_2019} to examine the variability of \rxj{}. We see that the system exhibits large amplitude variability by a factor of $\sim$10 over the approximately 17 years represented in Figure \ref{fig:flux_lc}. 
Two sources of variability are present in the light curves, first the intrinsic variability of the quasar and second the microlensing variability as the background source moving across magnification maps from quasar microlensing.
The intrinsic and microlensing variability can be separated by analyzing the flux ratio light curves between images shifted by appropriate time-delays, since the microlensing variability is uncorrelated between images.
Because of the large angular resolutions of \nustar, \swift, and \xmm, we can only measure the total source flux from all images from these observations.  
However, microlensing variability is clearly seen in the image resolved light curves from \textit{Chandra}, where image flux ratio changes were changing frequently and even reversing in brightness order \citep{Chartas_2019}. 
The variability amplitude monitored by \swift\ of a factor of $\sim$10 (lower panel of Figure~\ref{fig:flux_lc}) is comparable to that of previous Chandra monitoring period of a factor of $\sim$10 as well (upper panel of Figure~\ref{fig:flux_lc}), indicating microlensing is contributing similarly in the period monitored by \nustar, \swift, and \xmm.
This is consistent with the expectation that microlensing is ``constantly'' present in microlensing active systems, and it will take several decades for the source to move across the active microlensing regions in the microlensing magnification map \citep{Mosquera_2011}.

We compare the hard (10.0--80.0~keV) and soft (0.5--8.0~keV) band variability of \rxj. 
In the period where both hard and soft light curves are available between 5 June 2018 and 29 November 2020, where the hard band flux is measured by \nustar\ and soft by \xmm\ and \swift, we measure a factor of 3.7 change in the hard band and 5.5 in the soft band, showing smaller hard X-ray variability.
We next calculated the fractional root mean square (rms) amplitude $F_{var}$ \citep{Vaughan_2003} of the hard and soft band light curves, yielding $0.40 \pm 0.05$  and $0.57 \pm 0.02$ for the hard and soft bands respectively, where the soft band $F_{var}$ is larger by 3.2$\sigma$.

The total hard and soft X-ray flux is a combination of direct and reflected emission components.  Our spectral analysis of 7 joint hard and soft X-ray spectra has constrained the average reflection scales as $r_{refl} = 0.26$, and the evolution of the reflection fraction is plotted in Figure~\ref{fig:refl_curve} (left).  
We evaluate the variability of the reflection fraction over the seven epochs by fitting a constant model, yielding a residual  $\chi^2 = 7.6$ with 6 degrees of freedom, consistent with a constant model with 26\% probability.
For the last data point, there is an indication of an increase of the reflection fraction. 
The average reflection fraction for the first six observations is $r_{refl} = 0.20$, and the reflection fraction in the last observation, $ 0.65 ^{+ 0.22 }_{ -0.2 }$, is 
 2.3$\sigma$ away from this mean. 
We decompose the total X-ray flux into direct and reprocessed fluxes using the reflection fraction measurements for the seven epochs (Figure~\ref{fig:refl_curve}, right).
The total, direct, and reprocessed emission curves change by factors of $4.1 \pm 0.8$, $10.0 \pm 5.8$, and $3.4 \pm 3.2$ respectively, by comparing the highest over lowest flux measurements.  The large uncertainties for the direct and reprocessed components are caused by the measurement uncertainties of the reflection fraction. 
Examining the reprocessed emission curve shows that the excess variance is dominated by measurement errors, so the fractional variance is undefined, and a constant model fit to the reprocessed flux gives a residual $\chi^2$ of 3.47, or a 75\% probability for the constant model. 
However, if we discard the measurement uncertainties, the fractional variance of the flux estimates is 0.56 and 0.34 for the direct and reprocessed emission, respectively, suggesting a smaller variability amplitude for the reprocessed emission. 

Assuming quasar microlensing is contributing significantly to the total variability, the smaller variability amplitude and fractional rms in the hard band measured by \nustar\ indicate a larger hard X-ray emitting region than soft X-ray emitting region. 
However, variability amplitude and fractional rms are only moderately smaller, qualitatively suggesting the hard X-ray emission region is only moderately larger than the soft size of 10$r_g$ measured by \citet{Dai_2010}.
Although statistically insignificant, the reprocessed light curve suggests smaller variability amplitude as well.
This is consistent with a picture that the reflection is occurring in a region in the accretion disk with a range larger than the soft X-ray size in \rxj.  Remote reflection in the outer region of the accretion disk or torus as the majority of the reflection region in \rxj\ can still be viable because of large measurement uncertainties. Our result can be consistent with remote reflection contributing to a fraction of the reprocessed flux, e.g. 10--20\% as suggested by other studies \citep{2014ApJ...788...76W, 2017ApJ...836....2Z}.
We defer the quantitative quasar microlensing analysis to a future paper with the on-going extended monitoring data.

We have constrained the cutoff energy as 96$^{+47}_{-24}$~keV in the neutral reflection model, consistent with the survey results in \citet{Ricci_2018} which show characteristic cutoff energies for AGN of the order $10^2$ keV.  The ionized model yields similar cutoff energy values albeit with larger uncertainties in the upper bound.
We place our measurement in context with other quasars in the compactness-temperature diagram. As photon-photon collisions occur in the corona at very high energies, the photons decay into electron-positron pairs and then further annihilate into photons. Therefore pair production can become a runaway process in a sufficiently hot and compact corona, and the mechanism can acts as a natural thermostat. This theory is demonstrated by many quasars sitting along the pair production balance line in the compactness versus temperature ($l-\theta$) diagram, which is translated into observables in the $L_X-E_{cut}$ diagram  \citep{Bertola_2022}.
We show \rxj{} in the $L_X-E_{cut}$ plane compared to other quasars collected by \citet{Bertola_2022} and references therein in Figure \ref{fig:Lx_Ecut}, where the X-ray luminosity has been corrected by a total magnification factor $\mu_{\text{AGN}} = 48.2$ \citep{Paraficz_2018}, and \rxj\ lies on the allowable parameter space without violating the electron pair production limit.
This lens model is based on the point-like (quasar) image position from the CO emission measured by ALMA, and the model is consistent with those based on HST or 2.2$\mu$m quasar positions. Since the cutoff energy has a relatively low energy of $\sim$100~keV, whether \rxj\ lies in the allowable parameter space is almost independent of the magnification correction value. 

\begin{acknowledgments}

We would like to thank G.\ Lanzuisi and E.\ Bertola for providing the data points in Figure 6 and helpful discussion, and we acknowledge the support from NASA grants 80NSSC20K0034, 80NSSC22K0488, and 80NSSC23K0081.
AZ is supported by NASA under award number 80GSFC21M0002.

\end{acknowledgments}

\bibliography{bib}{}
\bibliographystyle{aasjournal}

\end{document}